\documentstyle[12pt]{article}

\newfont{\ftsect}{cmbx12 scaled\magstep2} 
\newfont{\fttitle}{cmbx12 scaled\magstep2}
\newfont{\ftauthor}{cmbx12 scaled\magstep1}
\newfont{\ftabstract}{cmsl12}
\newfont{\texte}{cmr12 scaled\magstep1}
\newfont{\xbf}{cmbx12 scaled\magstep1}
\newfont{\xit}{cmti12 scaled\magstep1}
\newfont{\xsl}{cmsl12 scaled\magstep1}
\newfont{\xtt}{cmti12 scaled\magstep1}
\newfont{\xsmall}{cmr12}
\newfont{\smallit}{cmti12}
\renewcommand{\sl}{\xsl}
  
\renewcommand{\small}{\xsmall}

\makeatletter
\@addtoreset{equation}{section}
\makeatother

\renewcommand{\subsection}[1]{\vspace{7mm}

\addtocounter{subsection}{1}\noindent
{\bf \thesubsection \ \ #1}
\vspace{4mm} 

\noindent}

\renewcommand{\title}[1]{\null\vspace{28mm} 

\begin{center}{\fttitle{\bf #1}}\end{center} }
\renewcommand{\author}[1]{\vspace{1mm} 

\begin{center}{\ftauthor #1}\end{center} }
\newcommand{\address}[1]{\vspace{-5mm} 

\begin{center}{\small  #1}\end{center} }
\renewcommand{\abstract}[1]{\vspace{15mm} 

\noindent{{\smallit Abstract.} {\small #1}} 
\large\texte  
}

\def\ftoday{{\sl  \number\day \space\ifcase\month 
\or Janvier\or F\'evrier\or Mars\or avril\or Mai
\or Juin\or Juillet\or Ao\^ut\or Septembre\or Octobre
\or Novembre \or D\'ecembre\fi
\space  \number\year}}    

\newcommand{\sla}{\raise.15ex\hbox{$/$}\kern -.57em} 
\newcommand{\Sla}{\raise.15ex\hbox{$/$}\kern -.70em}

\newcommand{\complex}{{\kern .1em {\raise .47ex
\hbox {$\scriptscriptstyle |$}}
    \kern -.4em {\rm C}}}
\newcommand{\real}{{{\rm I} \kern -.19em {\rm R}}}
\newcommand{\rational}{{\kern .1em {\raise .47ex
\hbox{$\scripscriptstyle |$}}
    \kern -.35em {\rm Q}}}
\renewcommand{\natural}{{\vrule height 1.6ex width
.05em depth 0ex \kern -.35em {\rm N}}}

\newcommand{\twiddle}{\lower.9ex\rlap{$\kern -.1em\scriptstyle\sim$}}

\newcommand{\eq}{\begin{equation}}
\newcommand{\eqn}[1]{\label{#1}\end{equation}}
\newcommand{\eea}{\end{eqnarray}}
\newcommand{\eqa}{\begin{eqnarray}}
\newcommand{\eqan}[1]{\label{#1}\end{eqnarray}}
\newcommand{\ba}{\begin{array}}
\newcommand{\ea}{\end{array}}
\newcommand{\eqac}{\begin{equation}\begin{array}{rcl}}
\newcommand{\eqacn}[1]{\end{array}\label{#1}\end{equation}}

\newcommand{\remarks}{\bigskip 

   \noindent{\bf Remarks:} \begin{enumerate}}
\newcommand{\skramer}{\end{enumerate}}

\begin{document}

  \rightline{\bf LYCEN~9749}
  \rightline{November 1997} 

\title{SYMMETRIES IN NUCLEAR, ATOMIC AND MOLECULAR 
SPECTROSCOPY\footnote{\small Lecture presented at the fifth 
{\it S\'eminaire Rhodanien de Physique} 
`Sur les sym\'etries en physique' held at the 
Domaine de Buffi\`eres 
(Dolomieu, France, 17~--~21 March 1997). 
To be published in {\bf Symmetries in Physics}, 
eds.~F. Gieres, M. Kibler, C. Lucchesi and O. Piguet 
(Editions Fronti\`eres, Paris, 1998). An extended version of this paper is
under preparation. The interested reader may contact the author for 
any comment, criticism and/or suggestion.} 
}

\begin{center}
dedicated to the memory of Alain Partensky 
\end{center}

\author{Maurice Kibler}
\address{Institut de Physique Nucl\'eaire de Lyon, 
         IN2P3-CNRS et Universit\'e Claude Bernard \\
         43 boulevard du 11 novembre 1918, 
         F-69622 Villeurbanne Cedex, France}
\abstract{Contents~: 
How to introduce groups and chains of groups in nuclear, atomic and 
molecular physics~? Some tools for connecting group theory and 
quantum mechanics (Wigner's theorem, Wigner-Eckart's theorem, 
Racah's lemma). Philosophy of some 
 qualitative and quantitative applications to spectroscopy.} 
\bigskip 

\vfill\eject

\title{SYMMETRIES IN NUCLEAR, ATOMIC AND MOLECULAR 
SPECTROSCOPY 
}

\begin{center}
dedicated to the memory of Alain Partensky 
\end{center}

\author{Maurice Kibler}
\address{Institut de Physique Nucl\'eaire de Lyon, 
         IN2P3-CNRS et Universit\'e Claude Bernard \\
         43 boulevard du 11 novembre 1918, 
         F-69622 Villeurbanne Cedex, France}
\abstract{Contents~: 
How to introduce groups and chains of groups in nuclear, atomic and 
molecular physics~? Some tools for connecting group theory and 
quantum mechanics (Wigner's theorem, Wigner-Eckart's theorem, 
Racah's lemma). Philosophy of some 
 qualitative and quantitative applications to spectroscopy.} 
\bigskip 

\noindent {\bf 1~~Introduction}
\medskip

This lecture is not a long and complete development on the use of symmetries 
(via group theory) in nuclear, atomic and molecular physics. It rather
addresses the three following questions. 

{\bf 1}. How the structures of group and of chain of groups enter nuclear, 
atomic and molecular spectroscopy~? 

{\bf 2}. How these structures can be exploited, in a quantum-mechanical 
framework, 
in the problems of state labelling and (external\footnote{\small We are not 
concerned here with the concept of spontaneous symmetry breaking which is
familiar 
in gauge theories of elementary particle physics. Let us however mention 
that this concept also occurs in nuclear physics (e.g., transition spherical
nucleus $\to$ deformed nucleus) and in molecular physics (e.g., Jahn-Teller
effect).}) symmetry breaking~?

{\bf 3}. How it is possible to associate a Wigner-Racah algebra to a group or 
a chain of groups for making easier 
the calculation of quantum-mechanical matrix elements~? 

We shall be mainly concerned in this lecture with the use of symmetries, via
group theory, for classifying wavefunctions and interactions and for
calculating matrix elements. The group theory to be used is along the lines
developed by G.~Frobenius, I.~Schur, S.~Lie, E.~Cartan, H.~Weyl, E.P.~Wigner
and G.~Racah. In this respect, the reader should note how Wigner's and Racah's
legacies are important for the applications of symmetries in nuclear, atomic, 
molecular and condensed matter physics. We shall not consider here external
(i.e., Lorentzian) symmetries and internal (i.e., gauge) symmetries which are 
of paramount importance in elementary particle physics. The latter symmetries 
shall be investigated in other lectures. It is enough to underline here the
importance of the gauge symmetry principle for generating 
(electroweak and strong) interactions and of the extended space-time symmetries 
for understanding the unification of external and internal symmetries through 
supersymmetry. 

Generally speaking, the  emphasis  in this lecture is on the philosophy rather
than on detailed calculations. However, four basic theorems as well as the
Wigner-Racah algebra for a finite or compact group are presented in a linear
and pedagogical way. 
Numerous examples illustrate the general ideas and a (necessarily
incomplete) list of applications closes this lecture. 
\bigskip

\noindent {\bf 2~~Introducing Groups and Chains of Groups}
\medskip

The introduction of the structure of group in connection with the concept of
conservation laws is familiar to the physicist. If a classical or quantum
physical system presents symmetries, then it is often 
possible to associate a Lie group, say $G$, to this system. 
This group describes some invariance of the physical system. As a result, there
exist $r$ conserved quantities or charges if $r$ is the order, 
i.e., the number of essential parameters, 
of the
invariance group $G$ (cf., Noether's theorem). In short, we have the sequence
$$
{\rm symmetries} \to {\rm group} \to {\rm invariance} \to {\rm conservation} \ 
{\rm laws}
$$
that is of special relevance for gauge theories and that will be discussed
in some of the other lectures. 

In spectroscopy, groups may be introduced in two complementary
ways (viz., as symmetry groups and as classification groups) 
that we shall discuss in turn with examples. Most of the examples shall be
worked out again in the other sections. 

{\bf 1}. {\em As symmetry groups}. If a Hamiltonien $H$ 
exhibits symmetries, we can introduce a 
(discrete or continuous) symmetry group $G$. 
The association of a group to a Hamiltonian is often a transcription 
of Curie's principle in its familiar formulation~: 
`The effects have the same symmetries as the causes'. The group $G$ 
is in general a subgroup of the
most general group that leaves the Hamiltonian invariant. According to Wigner's
theorem (to be proven in Section 3.1), 
it is then possible to classify the state vectors of 
$H$ and the interactions which occur besides $H$ by means of the irreducible
representations of $G$. A {\em symmetry} or {\em invariance} group is thus a 
{\em classification} group too. The appropriate sequence is now
$$
{\rm symmetries} \to {\rm group} \to {\rm invariance} 
                 \to {\rm conserved} \ {\rm quantum} \ {\rm numbers}
$$
where the notion of `conserved quantum numbers' or `good quantum numbers' 
arises from the fact that the Hamiltonian $H$, invariant under $G$, cannot
connect state vectors labelled by different irreducible representations of $G$ 
(this result shall be proven in Section 4.7). 

{\bf Example 1}. The Hamiltonian for a $N$-body system comprising 
$N$ identical 
particles is invariant under the symmetric group $S_N$. The wavefunctions for
this system have thus well-defined transformation properties under the
operations of $S_N$. We know that Physics selects the antisymmetric and the
symmetric (with respect to the permutation of two particles) wavefunctions for
fermions and bosons, respectively. 

{\bf Example 2}. The geometrical symmetry group of a (nonrelativistic)
Hamiltonian $H=T+V$ is commonly the discrete or continuous group that 
leaves the potential $V$ invariant. 

For instance, a complex atom is
rotationally invariant and its geometrical symmetry group is the full rotation
group. This group is isomorphic to the orthogonal group 
O(3) and we say that the geometrical
symmetry group of the atom is O(3). The invariance of ($V$ and) $H$ under O(3) 
ensures that the eigenfunctions of $H$ behave in a well-defined way under the
elements of O(3).   

As a second illustration, 
an ion embedded in a molecular environment and 
its environment possess a finite symmetry. This is the case for the complex ion
Ti(H$_2$O)$^{3+}_6$ for which the six dipoles H$_2$O are located at the vertices
of a (slightly deformed) octahedra and the central cation Ti$^{3+}$ occupies
the center of the octahedra. This complex ion 
is thus (almost) invariant under the octahedral group $O_h$. The same
invariance applies to the potential in the Hamiltonian $H$ of the ion 
Ti$^{3+}$~; in first approximation,  the geometrical symmetry group of 
$H$ is $O_h$. The invariance of $H$ under $O_h$ yields to molecular orbitals 
for Ti(H$_2$O)$^{3+}_6$ with well-defined properties with respect to the
elements of $O_h$. 

{\bf Example 3}. A symmetry or invariance group of a Hamiltonian often appears 
as a direct product of groups. For instance, if we want to take into
consideration both permutation and rotation symmetries in a complex atom with
$N$ electrons, we must combine $S_N$ and O(3) into the direct product
$S_N\otimes$O(3). Furthermore, if we make the approximation that the electronic
correlation is {\em negligible}, 
we can consider that $S_N \otimes$O(3)$^{\otimes N}$ 
is an {\em approximate} invariance group. Then, we foresee that the chain   
$S_N \otimes {\rm O(3)}^{\otimes N} \supset S_N \otimes {\rm O(3)}$ plays an
important r\^ole for a $N$-electron atom. 

{\bf Example 4}. In the case of the simplest atom, i.e., the hydrogen atom, the
group O(3) is not sufficient for a complete characterization of the
wavefunctions. For the discrete spectrum, there exists a larger group,
namely SO(4), whose generators commute with the Hamiltonian $H$ of the hydrogen
atom. This group is an invariance group 
(its generators commute with $H$ as a whole rather than 
with the kinetic and potential parts of $H$ separately). 
Thus, the relevant chain of groups to be
considered for the discrete spectrum is ${\rm O}(4) \supset {\rm O}(3)$. Such
a chain makes it possible to completely characterize or classify the discrete
eigenvectors of $H$ and to explain their degeneracy. 

{\bf 2}. {\em As classification groups}. On the other hand, we can use a single 
group or a chain
of groups for labelling the eigenvectors of a Hamiltonian $H$. Here, the 
single group or some groups of the chain do not correspond to
symmetries of $H$. The interest of such groups is to be found in the fact that
they allow to classify or label state vectors (or wavefunctions) and
interactions. 

{\bf Example 5}. Let us consider an ion with an electronic
configuration $nf^3$, $nf = 4f$ (lanthanide ion)  or  $5f$ (actinide ion). 
(The ion has several closed shells plus 
an $nf$ shell that is partly-filled with $3$ electrons.) 
The corresponding Hamiltonian involving kinetic and Coulomb interactions is
invariant under O(3). This invariance allows one to partially classify the 
$C^{N}_{2 (2 \ell + 1)} = 14!/(3! 11!) = 364$ wavefunctions for the 
configuration $nf^3$ ($\ell =3$ and $N=3$) with 
the angular momentum quantum numbers $L$ and $M_L$. We can also use 
the spin             quantum numbers $S$ and $M_S$ for labelling the
wavefunctions. Thus, the spectral group SU(2) that labels the spin may be
introduced in addition to the geometrical symmetry group O(3). However, the
group SU(2)$\otimes$O(3) is not sufficient for  labelling  completely the 364
state vectors $|nf^3 \tau S M_S L M_L )$ of the configuration $nf^3$. An
additional label $\tau$ is needed when two terms $^{2 S + 1}L$ with the same
$S$ and the same $L$ are permitted. For instance, there are two terms $^2H$
($S=1/2$ and $L=5$) in $nf^3$ and we need two labels $\tau_1$ and $\tau_2$ for
distinguishing them. It was shown by Racah~\cite{racah42} that this state 
labelling problem can be solved through the introduction of the chain of 
classification groups 
${\rm U}(7) \supset {\rm SO}(7) \supset G_2$, where $G_2$ is one of the
five exceptional groups of Cartan. The group U(7) corresponds to the fact that
any unitary transformation on the $2 \ell + 1 = 7$ orbitals for the shell $nf$
does not change the physics of the problem. The restriction U(7)$\to$SO(7)
indicates that we can consider real orbitals. Finally, the occurrence  of $G_2$
between SO(7) and SO(3) is a mathematical fact. As a result, the label $\tau$
can be replaced by $(w_1w_2w_3)(u_1u_2)$, where the triplet  $(w_1w_2w_3)$ 
stands     for an irreducible representation of the group SO(7) (of rank 3) and 
the doublet 
$(u_1u_2)$ for an irreducible representation of the group $G_2$ (of rank 2). 
Furthermore, the labelling afforded by the group U(7) is equivalent to the one
given by $S$ and $M_S$, so that it is not anymore necessary to consider the
spectral group SU(2) for the spin quantum numbers. The chain 
${\rm U}(7) \supset {\rm SO}(7) \supset G_2 \supset {\rm SO}(3)$ 
is thus appropriate for a $nf^3$ ion. The group
SO(3), or O(3) if we want to describe the parity of the wavefunctions,  is an
invariance group (and a classification group too) and the remaining groups of
the chain are classification groups only. This chain is sufficient for
characterizing completely the 
17 spectral terms $^{2 S + 1}L$ of the configuration
$nf^3$ and their associated state vectors  
$|nf^3 (w_1w_2w_3)(u_1u_2) S M_S L M_L )$. For example, the two terms $^2H$ 
correspond to $\tau_1 = (210)(11)$ and 
              $\tau_2 = (210)(21)$. We close by mentioning that the chain 
${\rm U}(7) \supset {\rm SO}(7) \supset G_2 \supset {\rm SO}(3)$ is also
extremely useful for classifying interactions like the Coulomb interactions
between the three equivalent electrons of $nf^3$. 
The classification of both state
vectors and interactions by means of irreducible representations of the groups
of a chain is essential for an easy and systematic calculation of matrix
elements. 

{\bf Example 6}. We continue with the ion of Example 5. We now take into 
account the spin-orbit interaction and we 
introduce the ion in a molecular or
crystal environment with $D_3$ (trigonal) symmetry. 
Then, it is interesting 
to replace the preceding  state vectors  by state vectors of the type
$|nf^3 (w_1w_2w_3)(u_1u_2) S L J a \Gamma \gamma )$. This amounts to complement
the chain of Example 5 by the chain ${\rm SU}(2) \supset D_3^{\star}$, where 
SU(2) describes the total angular momentum $J$ and the spinor group
$D_3^{\star}$ of $D_3$ labels the levels of the ion in its
environment. The label $a$ is to be used when the irreducible representation 
$(J)$ of SU(2) contains several times the irreducible representation $\Gamma$ 
of $D_3^{\star}$ and the label $\gamma$ distinguishes the various vectors
transforming as $\Gamma$. It is interesting to note that the actual 
symmetry group is $D_3$ and that all the other groups are classification
groups only. Note also that the labels $a$ and $\gamma$ are multiplicity 
labels without group-theoretical meaning. 

{\bf Example 7}. The problem of classifying state vectors can be seen as a
problem of finding a complete set of commuting operators. This may be
understood with the example of the Wigner-Hund SU(4) model of nuclei. 
The latter model combines 
the group SU(2)$_T$, which describes the Heisenberg isospin symmetry,  
    with 
the group SU(2)$_S$, for the spin of the nucleons. The resulting group 
SU(2)$_T \otimes $SU(2)$_S$ can be embedded in SU(4). The SU(4) symmetry is an
approximate symmetry for the nucleon-nucleon forces (it is broken by 
the Coulomb interaction betwen protons and the spin-orbit interaction for the
nucleons). The convenient chain
is then ${\rm SU}(4) \supset {\rm SU}(2)_T \otimes {\rm SU}(2)_S$ and the
corresponding state vectors read $| (p,q,r) \tau T M_T S M_S )$, where 
$T M_T$ and $S M_S$ refer to the group SU(2)$_T$ and SU(2)$_S$, respectively, 
and $(p,q,r)$ stands for an irreducible representation of SU(4). The 
labels $p$, $q$, $r$, $T$, $M_T$, $S$, and $M_S$ are not sufficient
in general for a complete labelling of the state vectors. A further label
$\tau$ is necessary. It is difficult to find a group-theoretical significance
of this `missing' label. However, it can be completely characterized by the
eigenvalues $\omega$ and $\varphi$ of two independent operators which commute
with the three Casimir operators of SU(4), the two isospin operators $T^2$ and
$T_3$, and the two spin operators $S^2$ and $S_3$. This leads to state vectors 
of the form $| (p,q,r) \omega \varphi T M_T S M_S )$ which are common
eigenvectors of nine commuting operators. 

We now examine in a more quantitative way how to exploit the 
introduction of groups and chains of groups by establishing 
two links between group theory and quantum mechanics.
\bigskip

\noindent {\bf 3~~Connecting Group Theory and Quantum Mechanics}
\medskip

{\bf 3.1 The Wigner theorem}
\smallskip

Let us consider a Hamiltonian $H$, defined on some Hilbert space ${\cal E}$, 
invariant under a finite or compact group $G$. We represent each element $R$ 
of $G$ by a linear operator $U_R$ that acts on ${\cal E}$. The invariance of 
$H$ under $G$ means that  
$$
\forall R \in G \qquad U_R^{-1} H U_R = H 
$$
(In other words, $H$ commutes with the group, isomorphic to $G$, spanned 
by the set $\{ U_R : R \in G \}$.) Let $E_{\lambda}$ be an eigenvalue of $H$ of
degeneracy $d$. It thus exists $d$ vectors $\phi_{\lambda j}$ in ${\cal E}$ 
such that    
$$
 H \phi_{\lambda j} = E_{\lambda} \phi_{\lambda j} \qquad j = 1, 2, \cdots, d
$$
We then have the series of trivial calculations 
$$
H (U_R \phi_{\lambda j}) =
(H U_R) \phi_{\lambda j} =
(U_R H) \phi_{\lambda j} =
U_R (H \phi_{\lambda j}) =
U_R (E_{\lambda} \phi_{\lambda j}) =
E_{\lambda} (U_R \phi_{\lambda j}) 
$$
As a result, the vector $U_R \phi_{\lambda j}$ is an eigenvector of $H$ with
the eigenvalue $E_{\lambda}$. Therefore, we can write $U_R \phi_{\lambda j}$ 
as a linear combination of the vectors $\phi_{\lambda k}$ with 
$k = 1, 2, \cdots, d$. Let us put 
$$
U_R \phi_{\lambda j} = \sum_{k=1}^{d} \phi_{\lambda k} D(R)_{kj} 
$$
where the coefficients of the linear combination, which depend on $R$ as well
as on $k$ and $j$, are denoted as $D(R)_{kj}$. The coefficients $D(R)_{kj}$ 
(for $k$ and $j = 1, 2, \cdots, d$) define a matrix $D(R)$. It is
straightforward to verify that   
$$
\forall R \in G \qquad 
\forall S \in G \qquad D(R)D(S) = D(RS) 
$$
so that $ D= \{ D(R) : R \in G \} $ constitutes a $d$-dimensional 
representation of $G$. This result may be summarized by the following 
theorem~\cite{wigner27}.  

{\bf Theorem 1} (Wigner's theorem). The eigenvectors corresponding to a given  
eigenvalue of a Hermitean\footnote{\small The
Hermitean operator $H$ may be replaced by  a  normal operator.} 
operator $H$ invariant under a finite or compact
group $G$ form a basis for a linear representation $D$ of $G$.

Since the group $G$ is finite or compact, there are two possibilities for $D$~:
The representation $D$ is either {\em irreducible} or {\em 
completely reducible}. We then have two definitions. 

{\bf Definition 1}. If the representation $D$ is irreducible, the degeneracy of
the $d$ functions $\phi_{\lambda 1}$, 
                  $\phi_{\lambda 2}$, $\cdots$, 
                  $\phi_{\lambda d}$ is said to be `essential' or `natural' or
`normal' with respect to the group $G$. 

{\bf Definition 2}. If the representation $D$ is (completely) reducible, the 
$d$ functions $\phi_{\lambda 1}$, 
              $\phi_{\lambda 2}$, $\cdots$, 
              $\phi_{\lambda d}$ are said to present an `accidental' degeneracy 
with respect to the group $G$. 

The Wigner theorem offers the possibility to classify wavefunctions and energy
levels of a Hamiltonian invariant under a group $G$ in terms of the
irreducible representations of $G$. These irreducible representations
constitute good quantum numbers for $H$. They are conserved in a sense to be
explained in Section 4.7. Before giving examples, we conclude that a 
{\em symmetry} or {\em invariance} group is also a {\em classification} group 
(the converse may not be true). 

{\bf Example 8}. Let 
$$
H = -\frac{1}{2} \frac{d^2}{dx^2} + \frac{1}{2} \omega^2 x^2 \qquad \omega > 0 
$$
be the Hamiltonian for a one-dimensional harmonic oscillator. Obviously, the
operator $H$ is invariant under the finite group $S_2$ 
($S_2 \sim Z_2 \sim C_2$). This group possesses two elements $E$ and $I$ 
corresponding to ${\rm E} : x \mapsto  x$ and 
                 ${\rm I} : x \mapsto -x$. Thus, it has two
irreducible representations $g$ (gerade) and $u$ (ungerade) that can be written 
$g = (1, 1)$ and $u = (1,-1)$ in the class space (E,I). (Other notations 
employed in spectroscopy  
for $g$ and $u$ are $[2  ]$ or $\Gamma_1$ or $A$ and 
                    $[1,1]$ or $\Gamma_2$ or $B$, respectively.) 
It is well-known that the spectrum of $H$ (energy levels $E_n$ and 
wavefunctions $\phi_n$) is given by 
$$
E_n = (n + \frac{1}{2}) \omega \qquad 
\phi_n (x) \sim H_n( \sqrt{\omega} x) {\rm exp}(-\frac{1}{2} \omega x^2) 
\qquad {\rm with} 
\qquad n \in {\bf N}  
$$
($H_n$ is the Hermite polynomial of degree $n$.)
Here, $d=1$ for each level $E_n$. Consequently, there is no degeneracy 
and each wavefunction $\phi_n$ must span a one-dimensional representation 
of the group $S_2$. This is indeed the case because the wavefunctions $\phi_n$
are symmetric or antisymmetric. The spectrum of $H$ is labelled by the 
  irreducible
representations of $S_2$~: $\phi_n$ with $n$ even spans the 
representation $g$ and     $\phi_n$ with $n$ odd        the 
representation $u$.  

{\bf Example 9}. The Hamiltonian $H$ for a nonrelativistic 
hydrogenlike atom reads 
$$
H = - \frac{1}{2} \Delta - Z \frac{1}{r} \qquad Z > 0 
\eqno (0)
$$
(The hydrogen atom corresponds to $Z=1$.)
The discrete energy spectrum is given by 
$$
E_n = \frac{1}{n^2} E_1 \qquad E_1 = - \frac{1}{2} Z^2 \qquad 
n \in {\bf N}^{*}
$$
The degeneracy degree for the level $E_n$ is 
$n^2 = \sum_{\ell = 0}^{n-1} (2 \ell + 1)$. It corresponds to the $n^2$
wavefunctions  $\phi_{n \ell m}$  with 
$ m = - \ell, - \ell + 1, \cdots, \ell $ and 
$ \ell = 0, 1, \cdots, n-1 $ associated to $E_n$. The kinetic energy 
$T= - (1/2) \Delta$ and the potential energy $ V = - Z / r $ are
separately invariant under the proper rotation group $R(3)$ in three dimensions 
(isomorphic 
to SO(3)). Therefore, $H$ is invariant under SO(3) too (and even under O(3)).  
The group SO(3) has a countable infinite number of irreducible 
representations $(\ell)$ with $\ell \in {\bf N}$. The irreducible
representation  $(\ell)$ is of dimension $2 \ell + 1$. 
The representation $D$, arising from Wigner's theorem, associated to $E_n$ is
$$
D \equiv D_{E_n} = \bigoplus_{\ell = 0}^{n-1} (\ell)
$$
which is reducible for $n \ne 1$. As a consequence, the group SO(3) does not
explain completely the degeneracy for $E_n$ when $n \ne 1$. The discrete
spectrum of $H$ exhibits accidental degeneracies with respect to the 
{\em geometrical symmetry group} SO(3) except for $n=1$. For $n$ and $\ell$
fixed, the group SO(3) explains the degeneracy of the $2 \ell + 1$ 
eigenvectors $ \{ \phi_{n \ell m} : m = - \ell, - \ell + 1, \cdots, \ell \} $. 
However, it does not explain the degeneracy of 
eigenvectors corresponding to a given value of $n$ and having different values
of $\ell$. It is possible (see the appendix) to show that  $D_{E_n}$  
turns out to be equivalent 
to the irreducible representation $(j,j)$, with $ j = (n-1)/2 $, 
of the group  ${\rm SO}(4) \sim {{\rm SU}(2) \otimes {\rm SU}(2)}/{Z_2}$. 
(The irreducible representations of ${\rm SU}(2) \otimes {\rm SU}(2)$ are 
denoted
as $(j_1,j_2)$ with $2 j_i \in {\bf N}$ for $ i = 1,2 $.) The group SO(4) 
is called a {\em degeneracy group}. For $n$ fixed, the degeneracy of the 
$n^2$ eigenvectors $\phi_{n \ell m}$ is natural with respect to SO(4). This 
group completely explains the degeneracies for the discrete spectrum of $H$ 
in the sense that each discrete level is associated to an irreducible
representation of SO(4). As a conclusion, the chain of groups 
${\rm SO}(4) \supset {\rm SO}(3)$ for the discrete spectrum contains two
types of groups~: the symmetry group SO(3) which describes the geometrical
symmetries of $T$ and $V$ (and thus $H$) and the degeneracy group SO(4) 
which labels the eigenvalues of $H$. The classification 
group SO(4) is also a symmetry group or invariance group 
for $H$ in view of the fact that its generators commute with $H$. It describes
the symmetries of $T + V$ as a whole. 

Besides the compact group SO(4) for the discrete spectrum,
noncompact groups may be introduced for the rest of the spectrum. As a matter
of fact, 
the pseudo-orthogonal group SO(3,1) describes the continuous spectrum 
and the Euclidean group E(3) the zero-energy point of the spectrum 
(see the appendix). These two groups play the r\^ole of 
classification groups and invariance groups.  

The preceding results 
can be generalized for a hydrogenlike atom in $N$ dimensions. The
groups SO($N+1$), SO($N,1$) and E($N$) can be seen to be invariance
groups for the discrete, continuous and zero-energy spectra of the
$N$-dimensional Coulomb system, respectively. These three groups have to be
distinguished from the {\em noninvariance group} 
SO($N+1,2$). The latter noncompact group is not an
invariance group in the sense that not all of its generators commute with the 
Hamiltonian of the $N$-dimensional hydrogen atom. It is rather a 
{\em dynamical group} in the sense that one of its irreducible representations 
contains all the wavefunctions for the spectrum of the 
Coulomb system in $N$ dimensions and that some of its 
generators may connect subspaces associated to different 
eigenvalues of the spectrum. The ordinary case $N=3$, 
which corresponds to ${\rm SO}(4,2) \sim {\rm SU}(2,2)/Z_2$, 
is studied at length in the literature. 

{\bf Example 10}. The Hamiltonian $H$ for an isotropic 
            harmonic oscillator in $N=3$ dimensions is
$$
H = - \frac{1}{2} \Delta + \frac{1}{2} \omega^2 r^2 \qquad \omega >0 
$$
The spectrum of $H$ is entirely discrete. The energy levels are  
$$
E_n = ( n + \frac{3}{2} ) \omega \qquad n \in {\bf N} 
$$
The subspace 
${\cal E}_n = \{ \Psi_{n_1n_2n_3} : n_i \in {\bf N}, \ i=1,2,3 \ | \ 
                           n_1 + n_2 + n_3 = n \}$,
where the $\Psi$'s are simple products of the $\phi$'s of Example 8, 
is associated to the level $E_n$. 
Thus, the degeneracy degree for the level $E_n$ is 
$C^{n}_{n + N - 1} = (n+1)(n+2)/2$. The kinetic energy $T= - (1/2) \Delta$ and the potential 
energy $ V = (1/2) \omega^2 r^2 $ are
invariant under SO(3). Therefore, the group SO(3) is a 
{\em geometrical symmetry group} for $H$. According to Wigner's theorem, the
eigenfunctions in ${\cal E}_n$ span a representation $D \equiv D_{E_n}$. 
This representation is in general reducible. (Hint~: For $n=2$, we have 
dim${\cal E}_n = 6$ and it does not exist a true irreducible representation 
$(\ell)$ of SO(3) such that $2 \ell + 1 = 6$.) It is possible to show that 
the Hamiltonian $H$,  as considered as a whole  (by writing it in terms of
annihilation and creation boson operators), 
is invariant under the group SU(3).  It turns out that the wavefunctions of
${\cal E}_n$ generate the irreducible representation $(n,0)$ of SU(3). 
The group SU(3) is thus a {\em degeneracy group}. 
(The irreducible representations of SU(3) are characterized by couples 
$(p,q) \in {\bf N}^2$. 
The dimension of the representation $(p,q)$ 
is dim$(p,q) = (p+1)(q+1)(p+q+2)/2$.) 
For instance, we have the associations 

\indent
$ D_{E_1} = (0,0) = (0) : 1s$ shell              (dim${\cal E}_1 =  1)$ \\
\indent
$ D_{E_2} = (1,0) = (1) : 1p$ shell              (dim${\cal E}_2 =  3)$ \\ 
\indent
$ D_{E_3} = (2,0) = (0) \oplus (2) : 2s1d$ shell (dim${\cal E}_3 =  6)$ \\ 
\indent
$ D_{E_4} = (3,0) = (1) \oplus (3) : 2p1f$ shell (dim${\cal E}_4 = 10)$ 

\noindent 
where we have indicated the decompositions of the representations 
($n,0$) of SU(3) into representations ($\ell$) of SO(3) as well as the
corresponding nuclear shells (cf., the  Mayer-Jensen shell model and the 
Elliott  SU(3)  rotation model). 
We note that the levels $E_1$ and $E_2$ do not exhibit accidental degeneracy
with respect to SO(3) but that $E_3$ and $E_4$ do. The first nuclear magic
numbers $A = 2$, 8, and 20 correspond to the groupings $1s$, $1s + 1p$, and 
$1s + 1p + 2s1d$, respectively. 

The extension from the three-dimensional oscillator to the $N$-dimensional 
isotropic harmonic oscillator is immediate. In the  $N$-dimensional  case,
the geometrical symmetry group is SO(N) and 
the group SU(N) is a degeneracy (and thus invariance) group. 
The corresponding spectrum has accidental 
degeneracy with respect to SO(N). 
However, all the degeneracies are natural with respect to SU(N). 
It is to be noticed that the real noncompact 
symplectic group Sp(2$N,{\bf R}$) is a useful 
noninvariance group. The
chain  ${\rm Sp}(2N,{\bf R}) \supset {\rm SU}(N) \supset {\rm SO}(N)$ 
is of importance when looking for a dynamical group 
for  the  $N$-dimensional oscillator system. 

{\bf Example 11}. We close this series with the example of a 
three-dimensional nonrelativistic Hamiltonian $H = -(1/2) \Delta + V$ 
where V is a central potential with $V(r) \ne - Z/ r$ and 
$ V(r) \ne (1/2) \omega^2 r^2 $. The operators $V$ and $-(1/2) \Delta$ 
(and thus $H$) are invariant under  ${\rm SO}(3) \otimes C_i$,  where 
$C_i \sim S_2$. We know that the
discrete spectrum, if any, corresponds in general to energies of type 
$E_{n\ell}$. The level $E_{n \ell }$ is associated to 
${\cal E}_{n \ell} = \{ R_{n \ell} (r) Y_{\ell m} (\theta, \varphi) :  
m = -\ell, -\ell + 1, \cdots, \ell \}$, 
a subspace of $2 \ell + 1$ wavefunctions. Here, 
$\ell$ belongs to ${\bf N}$ and $n - \ell -1$ is the number of nodes
(excluding $0$ and $\infty$) of the radial wavefunction $R_{n \ell}$. 
The Wigner theorem can be invoked for both SO(3) and $C_i$. 
Let $D \equiv D_{E_{n \ell}}$ be the representation (of SO(3) and $C_i$) 
spanned by the subspace ${\cal E}_{n \ell}$. For the group SO(3), we have 
$D_{E_{n \ell}} \equiv (\ell)$ : 
The degeneracy for the energy level $E_{n \ell}$ is natural with respect
to the group SO(3). On the other hand, the behaviour under the group $C_i$ 
of the vectors of ${\cal E}_{n \ell}$ is trivial. The group $C_i$ consists
     of the elements ${\rm E} : {\vec r} \mapsto   {\vec r}$ and
                     ${\rm I} : {\vec r} \mapsto - {\vec r}$. Then, 
for I (which
corresponds to $\theta  \mapsto \pi     - \theta$ and
               $\varphi \mapsto \varphi + \pi   $), we have
$$
U_{\rm I} :              R_{n \ell} (r) Y_{\ell m} (\theta, \varphi)  \mapsto 
             (-1)^{\ell} R_{n \ell} (r) Y_{\ell m} (\theta, \varphi) 
$$
Therefore, the subspace 
${\cal E}_{n \ell}$ constitutes a basis for a reducible representation of  
$C_i$~: The degeneracy of the $2 \ell + 1$ wavefunctions of $E_{n \ell}$ 
is accidental with respect to the group $C_i$. The wavefunctions 
$  R_{n \ell} Y_{\ell m} $  may be labelled by irreducible representations of
$C_i$ since they are even or odd under $C_i$ according to as 
$\ell$ is even or odd. Therefore, the decomposition of 
$D_{E_{n \ell}}$ leads to $D_{E_{n \ell}} = (2 \ell + 1)g$ for $\ell$ even   
                       or $D_{E_{n \ell}} = (2 \ell + 1)u$ for $\ell$ odd.
\medskip

{\bf 3.2 The restriction group $\to$ subgroup} 
\smallskip 

The concept of a chain of groups $G_0 \supset G_1$ is important in physics
especially in connection with symmetry breaking mechanisms and/or 
perturbation theory. In the framework of perturbation theory, the group $G_0$
may be an invariance group for an unperturbed Hamiltonian $H_0$ while 
the subgroup $G_1$ of $G_0$  
may be an invariance group for a perturbed Hamiltonian $H_0 + H_1$. 
The passage from  $H_0$ to $H_0 + H_1$ thus corresponds to a symmetry breaking
where the symmetry of the Hamiltonian is lowered from $G_0$ to $G_1$. A basic
result concerning the restriction $G_0 \to G_1$ is given by the following
theorem. 

{\bf Theorem 2}. Let $D_0 = \{ D_0(R) : R \in G_0 \}$ be a 
linear representation, of dimension $d$, of a group $G_0$. 
The restriction
$$
D_1 = \{ D_1(R) = D_0(R) : R \in G_1 \subset G_0 \}
$$ 
of $D_0$ to a subgroup $G_1$ of the group $G_0$ furnishes 
a representation of $G_1$.

The proof easily follows from the fact that 
$\forall R \in G_1$ and $\forall S \in G_1$ 
we have $D_1(R) D_1(S) = D_1(RS)$. 
Therefore, every representation of $G_0$ yields a representation of $G_1$. 
If $D_0$ is a reducible representation of $G_0$, then $D_1$ is necessarily a
reducible representation of $G_1$. On the other hand, if 
$D_0$ is an irreducible representation of $G_0$, then the 
representation $D_1$ may be a reducible or irreducible 
representation of $G_1$.

Theorem 2 is very useful in the case of external symmetry breaking, i.e., in
the case where the geometrical symmetry of a system is lowered through some
external action as, for example, in the Zeeman effect and in the
(homogeneous or inhomogeneous\footnote{\small For instance, 
the inhomogeneous Stark effect 
arises when a partly-filled shell ion is embedded in a crystal~; such an ion is
thus subjected to an (inhomogeneous) crystalline electric field which has, 
according to Curie's principle, the
symmetry of the environment of the ion.}) Stark effect. The restriction 
$G_0 \to G_1$, in terms of irreducible representations of $G_0$ and $G_1$, 
is of central importance to
see how the energy levels of $H_0$ evolve when turning on the perturbation
$H_1$. This level splitting when passing from the symmetry $G_0$ to the lower
symmetry $G_1$ is formally obtained by looking at the decomposition of the 
irreducible representations of $G_0$ into a
direct sums of irreducible representations of $G_1$. 

{\bf Example 12}. Let us consider the case of a complex 
ion with configuration 4$f^1$, like the ion Ce$^{3+}$, 
embedded in a crystal environment of octahedral (or cubical) symmetry.  
If we do not consider 
the spin-orbit interaction for the $4f$ electron, the ground state for the 
configuration 4$f^1$ of the free ion is a term $^2F$ 
($S = s = \frac{1}{2}$, $L = \ell  = 3$). This term spans the irreducible
representation $(3_u)$ (of dimension $d=7$) of the group O(3), 
an invariance group for the free ion. (The irreducible representations of the 
direct product ${\rm O}(3) \sim {\rm SO}(3) \otimes C_i$ 
are of type ($\ell_g$)
or ($\ell_u$) with $\ell \in {\bf N}$.) By using standard methods, we can 
show that the decomposition of $(3_u)$ into irreducible representations 
of the complete octahedral group $O_h = {O} \otimes C_i$ leads to 
$$
(3_u) = A_{2u} \oplus T_{1u} \oplus T_{2u} 
$$
where $T_{1u}$ and $T_{2u}$ are two three-dimensional irreducible
representations of $O_h$ and $A_{2u}$ is a one-dimensional 
representation  of $O_h$. Therefore, the atomic term $^2F$ gives rise 
to three
(crystal-field or molecular) terms $^2A_{2u}$, $^2T_{1u}$ and $^2T_{2u}$ when
we pass from the free ion to the ion in its cubical surrounding.

{\bf Example 13}. To go further with Example 12, we can now deal with 
the case where we take into consideration the spin-orbit interaction 
for the $4f$ electron. The term  
$^2F$ then splits into two multiplets $^2F_{7/2}$ 
and $^2F_{5/2}$. The relevant group for the free ion is SU(2) 
(with ${\rm SO}(3) \sim {\rm SU}(2) / Z_2$) and the one for the ion 
in its cubical 
surrounding is the `doubled' or spinor group $O^{\star}$ 
(with $O \sim O^{\star} / Z_2$). The state vectors
for the multiplets $^2F_{7/2}$ and $^2F_{5/2}$ span the irreducible
representations (5/2) and (7/2) (of dimensions $d=6$ and $d=8$) of SU(2), 
respectively. The restriction ${\rm SU}(2) \to O^{\star}$ yields  
$$
( 5/2 ) = \Gamma_7 \oplus \Gamma_8 \qquad 
( 7/2 ) = \Gamma_6 \oplus \Gamma_7 \oplus \Gamma_8 
$$
so that the multiplets $^2F_{7/2}$ and $^2F_{5/2}$ are split 
(in the absence of accidental degeneracies) according 
to one doublet ($\Gamma_7$) plus one quadruplet ($\Gamma_8$) and 
two doublets   ($\Gamma_6$  and $\Gamma_7$) plus one quadruplet ($\Gamma_8$), 
respectively. (The irreducible representations of SU(2) 
are denoted as ($j$) with $2j \in {\bf N}$ while $\Gamma_6$, $\Gamma_7$ and 
$\Gamma_8$ are irreducible representations of $O^{\star}$.)
\bigskip

\noindent{\bf 4~~Wigner - Racah Algebra}
\medskip

An important task in spectroscopy is to calculate matrix elements in order 
to determine energy spectra and transition intensities. One way to 
incorporate symmetry considerations connected to a chain of groups
(involving symmetry groups and classification groups) is to use the 
`Wigner-Racah calculus' associated to the chain under consideration.   
The `Wigner-Racah calculus' or 
    `Wigner-Racah  algebra' associated to a group $G$ or a chain of groups 
$G \supset H$ is generally understood as the set of algebraic manipulations
concerning the coupling and recoupling coefficients for the group $G$. This
`algebra' may be also understood as a true algebra (in the mathematical 
sense)~: It is the (infinite dimensional) Lie algebra spanned by the
irreducible unit tensor operators or Wigner operators of the group $G$. We 
shall mainly focus here on the basic aspects of 
the `algebra' of the coupling and recoupling coefficients of $G$. 
The Wigner-Racah calculus was originally developed for simply-reducible 
(i.e., ambivalent\footnote{\small A group $G$ is said to be ambivalent if each
element of $G$ and its inverse belong to a same conjugation class.}
 plus multiplicity-free\footnote{\small A group $G$ is said to be 
multiplicity-free
if the Kronecker product of two arbitrary irreducible representations of $G$ 
contains at most once each irreducible representation of $G$.}) 
groups~\cite{wigner41,wigner65,wigner68}, for 
the rotation group~\cite{wigner65,racah42} and for some groups of interest in
molecular and condensed matter physics~\cite{griffith62,stk70,kibler7982}.  
The extension to an arbitrary finite or compact group can be achieved 
and we present in what follows the ingredients for such an extension (that is 
of great interest in nuclear, atomic, molecular, and condensed matter physics 
as well as in quantum chemistry).

\newpage 

{\bf 4.1 Preliminaries}
\smallskip

Let us consider an arbitrary finite or compact continuous group $G$ having 
the irreducible representation classes (IRC's) $a$, $b$, $\cdots$. The 
identity IRC, customarily noted $A_1$ or $\Gamma_1$ in molecular physics, 
is denoted by 0. To each IRC $a$, we associate a unitary matrix 
representation ${\cal D}^a$. Let [a] be the dimension of ${\cal D}^a$. The 
$\alpha$-$\alpha'$ matrix element of the representative ${\cal D}^a(R)$ for 
the element $R$ in $G$ is written ${\cal D}^a(R)_{\alpha\alpha'}$. (For $a=0$, 
we use $\alpha = \alpha' = 0$.) The sum $\chi^a(R) = \sum_{\alpha} 
{\cal D}^a(R)_{\alpha\alpha}$ stands for the character of $R$ in 
${\cal D}^a$. The ${\cal D}^a(R)_{\alpha\alpha'}$ and $\chi^a(R)$ satisfy 
orthogonality relations (e.g., the so-called great orthogonality theorem) 
that are very familiar to the physicist. Finally, note that 
$ \int_G \cdots dR $ identifies to $\sum_{R \in G} \cdots$ 
and that $\left\vert G \right\vert = \int_G dR$ corresponds to 
the order  of $G$ in the case where $G$ is a finite group or
the volume of $G$ in the case where $G$ is a continuous compact group. 
\medskip 

{\bf 4.2 Clebsch-Gordan coefficients}
\smallskip 

The direct product $a \otimes b$ of two IRC's $a$ and $b$ of $G$ can be in 
general decomposed into a direct sum of IRC's of $G$. This leads to the 
Clebsch-Gordan series
$$
a \otimes b = \bigoplus_c \sigma (c | a \otimes b) c
\eqno (1)
$$
where $\sigma (c | a \otimes b)$ denotes the number of times the IRC $c$
occurs in $a \otimes b$. The integers 
      $\sigma (c | a \otimes b)$ may be determined through the character formula
$$
       \sigma (c | a \otimes b) = 
\left\vert G \right\vert^{-1}{\int_G} \chi^c(R)^* \chi^a(R) \chi^b(R) dR
\eqno (2)
$$
In terms of matrix representations, Eq.~(1) reads
$$
{\cal D}^a \otimes {\cal D}^b \simeq 
                  \bigoplus_c \sigma (c | a \otimes b) {\cal D}^c
\eqno (3)
$$
Therefore, there exists a unitary matrix $U^{ab}$ such that
$$
(U^{ab})^\dagger {\cal D}^a(R) \otimes {\cal D}^b(R) U^{ab} = \bigoplus_c 
\sigma (c | a \otimes b) {\cal D}^c(R)
\eqno (4)
$$
$$
{\cal D}^a(R) \otimes {\cal D}^b(R) = \bigoplus_c \sigma (c | a \otimes b) 
U^{ab} {\cal D}^c(R) (U^{ab})^\dagger 
\eqno (5)
$$
for any $R$ in $G$. It is a simple exercice in linear algebra to transcribe 
(4) and (5) in matrix elements. We thus have
$$
\sum_{\alpha \beta \alpha' \beta'} (a b \alpha \beta \vert \rho c \gamma )^*
{\cal D}^a(R)_{\alpha \alpha'}
{\cal D}^b(R)_{\beta   \beta'}     (a b \alpha'\beta'\vert \rho'c'\gamma')
$$
$$
= \Delta (c \vert a \otimes b) \delta (\rho'\rho) \delta(c'c)
{\cal D}^c(R)_{\gamma \gamma'}
\eqno(6)
$$
and
$$
{\cal D}^a(R)_{\alpha \alpha'} {\cal D}^b(R)_{\beta   \beta'}
= \sum_{\rho c \gamma \gamma'} (a b \alpha \beta \vert \rho c \gamma )
{\cal D}^c(R)_{\gamma \gamma'} (a b \alpha'\beta'\vert \rho c \gamma')^* 
\eqno(7)
$$
for any $R$ in $G$. In Eqs.~(6) and (7), 
$(a b \alpha \beta \vert \rho c \gamma)$ stands for an element of the matrix 
$U^{ab}$~:
$$
 (a b \alpha \beta \vert \rho c \gamma) 
                           = \left( U^{ab} \right)_{\alpha \beta,\rho c \gamma}
\eqno(8)
$$
Each row index of $U^{ab}$ consists of two labels ($\alpha$ and $\beta$)
according to the rules of the direct product of two matrices. This is 
the same thing for each column index of $U^{ab}$~, 
i.e., two labels ($c$ and $\gamma$)
are required. However, when $c$ appears several times in $a \otimes b$, a third
label (the multiplicity label $\rho$) is necessary besides $c$ and $\gamma$.
Hence, the summation over $\rho$ in (7) ranges from 1 to 
$\sigma (c \vert a \otimes b)$. Finally, in Eq.~(6), $\delta$ denotes the usual
Kronecker delta while $\Delta (c \vert a \otimes b) = 0$ or 1 according to 
whether as $c$ is contained or not in $a \otimes b$. 

The matrix elements $(a b \alpha \beta \vert \rho c \gamma)$ are termed 
Clebsch-Gordan coefficients (CGc's) or vector coupling coefficients. The 
present introduction clearly emphasizes that the CGc's of a group $G$ are 
nothing but the elements of the unitary matrix which reduces the direct product 
of two irreducible matrix representations of $G$. As a consequence, the CGc's 
satisfy two orthonormality relations associated to the unitary property of 
$U^{ab}$~:
$$
\sum_{\alpha \beta} (a b \alpha \beta \vert \rho c \gamma )^* 
                    (a b \alpha \beta \vert \rho'c'\gamma') = 
\Delta(c \vert a \otimes b) 
\delta(\rho' \rho) \delta(c' c) \delta(\gamma' \gamma)
\eqno(9)
$$
$$
\sum_{\rho c \gamma} (a b \alpha \beta \vert \rho c \gamma)
                     (a b \alpha'\beta'\vert \rho c \gamma)^* = 
\delta(\alpha' \alpha) \delta(\beta'  \beta )
\eqno(10)
$$
Note that (9) and (10) are conveniently recovered by specializing $R$ to the
unit element $E$ of $G$ in (6) and (7), respectively.

Equations (6) and (7) show that the CGc's are basis-dependent coefficients. In
this regard, it is important to realize that (6) and (7) are not sufficient to 
define unambiguously the CGc's of the group $G$ once its irreducible
representation matrices are known. As a matter of fact, the relation
$$
(a b \alpha \beta \vert r    c \gamma) = \sum_{\rho} 
(a b \alpha \beta \vert \rho c \gamma) M(a b,c)_{\rho r}
\eqno(11)
$$
where $M(ab,c)$ is an arbitrary unitary matrix, defines a new set of CGc's since
(6) and (7) are satisfied by making replacements of the type $\rho \to r$. 
The CGc's associated to a definite choice for the irreducible representation
matrices of $G$ are thus defined up to a unitary transformation, a fact that 
may be exploited to generate special properties of the CGc's.

Various relations involving elements of irreducible representation matrices and
CGc's can be derived from (6) and (7) by using the unitary property both
for the representation matrices and the Clebsch-Gordan matrices. For instance,
from (6) we obtain
$$
\sum_{\alpha' \beta'}{\cal D}^a(R)_{\alpha \alpha'}
                     {\cal D}^b(R)_{\beta  \beta '}
(a b \alpha' \beta' \vert \rho c \gamma') = \sum_{\gamma} 
                     {\cal D}^c(R)_{\gamma \gamma'}
(a b \alpha  \beta  \vert \rho c \gamma )
\eqno(12)
$$
$$
(a b \alpha' \beta' \vert \rho c \gamma') = \sum_{\alpha \beta \gamma}
(a b \alpha  \beta  \vert \rho c \gamma )
{\cal D}^a(R)_{\alpha \alpha'}^*
{\cal D}^b(R)_{\beta  \beta '}^*
{\cal D}^c(R)_{\gamma \gamma'}
\eqno(13)
$$
for any $R$ in $G$. In the situation where the elements of the irreducible
representation matrices of $G$ are known, (12) or (13) provides us with a
system of linear equations useful for the calculation of the CGc's of $G$. 

The combination of (7) with the great orthogonality theorem for $G$ yields
the integral relation
$$
\vert G \vert^{-1} \int_G 
{\cal D}^a(R)_{\alpha \alpha'}
{\cal D}^b(R)_{\beta  \beta '}
{\cal D}^c(R)_{\gamma \gamma'}^* dR
= [c]^{-1} \sum_{\rho} 
(a b \alpha  \beta  \vert \rho c \gamma) 
(a b \alpha' \beta' \vert \rho c \gamma')^*
\eqno(14)
$$
which also is useful for the calculation of the CGc's of $G$ in terms of the
elements of the irreducible representation matrices of $G$. Note that when
$a \otimes b$ is multiplicity-free (i.e., when there is no summation on 
$\rho$ in (14)), Eq.~(14) allows us to determine the 
$(a b \alpha \beta \vert c \gamma)$ for all $\alpha$, $\beta$ and $\gamma$ up 
to an arbitrary phase factor $h(ab,c)$~; more precisely, we then have
$$
(a b \alpha \beta | c \gamma) = {\rm e}^{{\rm i} h(ab,c)} 
\frac{
\int_G   {\cal D}^a(R)_{\alpha \alpha'} 
         {\cal D}^b(R)_{\beta  \beta '}
         {\cal D}^c(R)_{\gamma \gamma'}^* dR}
{\lbrace \frac{ |G| }{ [c] } 
\int_G   {\cal D}^a(R)_{\alpha' \alpha'}
         {\cal D}^b(R)_{\beta ' \beta '}
         {\cal D}^c(R)_{\gamma' \gamma'}^* dR \rbrace^{ \frac{1}{2} } }
\eqno (15)
$$

It appears from Eqs.~(12)-(15) that $c$ does not generally play the same 
r\^ole as
$a$ and $b$ in $(a b \alpha \beta \vert \rho c \gamma)$. Therefore, 
               $(a b \alpha \beta \vert \rho c \gamma)$ does not generally 
exhibit simple symmetry properties under permutation of $a$, $b$ and $c$. It 
is to be showed in the following how the CGc's may be symmetrized thanks to a 
2-$a \alpha$ symbol.
\medskip 

{\bf 4.3 The $2-a \alpha$ symbol}
\smallskip 

Let us define  the 2-$a \alpha$ symbol through
$$
\pmatrix{
a&b\cr
\alpha&\beta\cr
} = [a]^{\frac{1}{2}} (b a \beta \alpha \vert 00)
\eqno(16)
$$
The 2-$a \alpha$ symbol makes it possible to pass from a given irreducible
matrix representation to its complex conjugate. This is reflected by the two
relations
$$
\sum_{\alpha \alpha'} \pmatrix{
a&b\cr
\alpha&\beta\cr
}^*
{\cal D}^a(R)_{\alpha\alpha'} \pmatrix{
a&b'\cr
\alpha'&\beta'\cr
}
= \Delta (0 \vert a \otimes b) \delta (b' b) {\cal D}^b(R)_{\beta \beta '}^*
\eqno(17)
$$
$$
\sum_{\beta \beta'} \pmatrix{
a&b\cr
\alpha&\beta\cr
} 
{\cal D}^b(R)_{\beta\beta'}^* \pmatrix{
a'&b\cr
\alpha'&\beta'\cr
}^*
= \Delta (0 \vert a \otimes b) \delta (a' a) {\cal D}^a(R)_{\alpha\alpha'}
\eqno(18)
$$
that hold for any $R$ in $G$. The proof of (17) and (18) is delicate~; it 
starts with the introduction of (16) into the left-hand sides of (17)
and (18) and requires the successive use of (13), (7), (9) and (13), of
the great orthogonality theorem, and of (9). By taking $R = E$ in (17)
and (18), we get the useful relations
$$
\sum_{\alpha}\pmatrix{
a&b\cr
\alpha&\beta\cr
}^*
\pmatrix{
a&b'\cr
\alpha&\beta'\cr
}
= \Delta (0\vert a\otimes b) \delta(b'b) \delta(\beta'\beta)
\eqno(19)
$$
$$
\sum_{\beta}\pmatrix{
a&b\cr
\alpha&\beta\cr
}
\pmatrix{
a'&b\cr
\alpha'&\beta\cr
}^*
 = \Delta (0\vert a\otimes b) \delta(a'a) \delta(\alpha'\alpha)
\eqno(20)
$$

The  2-$a \alpha$ symbol turns out to be of relevance for handling phase
problems. In this regard, both (17) and (18) lead to
$$
\delta(a'a) \sum_{ \alpha\alpha' } \pmatrix{
a&a'\cr
\alpha&\alpha'\cr
}^*
\pmatrix{
a'&a\cr
\alpha'&\alpha\cr
}
= \Delta (0 \vert a \otimes a') [a] c_a
\eqno(21)
$$
where the Frobenius-Schur coefficient
$$
c_a = \vert G \vert^{-1} \int_G \chi^a (R^2) dR
\eqno(22)
$$
is 1, $-1$, or 0 according to as ${\cal D}^a$ is orthogonal, 
symplectic, or complex. The conjugating matrix to pass from 
${\cal D}^a$ to $( {\cal D}^a )^*$ satisfies 
$$
c_a \pmatrix{
a'&a\cr
\alpha'&\alpha\cr
}
= \delta(a'a)\pmatrix{
a&a'\cr
\alpha&\alpha'\cr
}
\eqno(23)
$$
(cf., the Frobenius-Schur theorem). 
\medskip 

{\bf 4.4 The $( 3-a \alpha )_{\rho}$ symbol}
\smallskip 

We now define the (3-$a \alpha)_\rho$ symbol via
$$
\pmatrix{
a&b&c\cr
\alpha&\beta&\gamma\cr
}_\rho
= \sum_{\rho' c' \gamma'} [c']^{-\frac{1}{2}} M(ba,c')_{\rho' \rho} \pmatrix{
c&c'\cr
\gamma&\gamma'\cr
}
(b a \beta \alpha \vert \rho' c' \gamma')
\eqno(24)
$$
where $M(ba,c')$ is an arbitrary unitary matrix. Conversely, each CGc can be
developed in terms of (3-$a \alpha)_\rho$ symbols since the inversion of 
(24) gives 
$$
(a b \alpha \beta \vert \rho c \gamma) = 
[c]^{\frac{1}{2}} \sum_{\rho' c' \gamma'}
M(ab,c)_{\rho\rho'}^*\pmatrix{
c'&c\cr
\gamma'&\gamma\cr
}^*
\pmatrix{
b&a&c'\cr
\beta&\alpha&\gamma'\cr
}_{\rho'}
\eqno(25)
$$

All the relations involving CGc's may be transcribed in function of
(3-$a\alpha)_\rho$ symbols. For example, the introduction of (25) into 
(6) and (7) yields after nontrivial calculations
$$
\sum_{\alpha \beta \alpha' \beta'} \pmatrix{
a&b&c\cr
\alpha&\beta&\gamma\cr
}^*_\rho
{\cal D}^a(R)_{\alpha \alpha'}
{\cal D}^b(R)_{\beta  \beta' }
\pmatrix{
a&b&c'\cr
\alpha'&\beta'&\gamma'\cr
}_{\rho'} 
$$
$$
= \Delta (0 \vert a \otimes b \otimes c) \delta(\rho' \rho) \delta(c' c) 
[c]^{-1} {\cal D}^c(R)_{\gamma\gamma'}^* 
\eqno(26)
$$
and
$$
{\cal D}^a(R)_{\alpha\alpha'}{\cal D}^b(R)_{\beta\beta'} = 
\sum_{\rho c \gamma \gamma'} [c] 
\pmatrix{
a&b&c\cr
\alpha&\beta&\gamma\cr
}_\rho
{\cal D}^c(R)_{\gamma\gamma'}^*\pmatrix{
a&b&c\cr
\alpha'&\beta'&\gamma'\cr
}_\rho^*
\eqno(27)
$$
for any $R$ in $G$. The orthogonality relations
$$
\sum_{\alpha \beta} \pmatrix{
a&b&c\cr
\alpha&\beta&\gamma\cr
}_\rho^*
\pmatrix{
a&b&c'\cr
\alpha&\beta&\gamma'\cr
}_{\rho'}
= \Delta (0 \vert a \otimes b \otimes c) \delta(\rho' \rho) \delta(c' c)
\delta(\gamma' \gamma) [c]^{-1}
\eqno(28)
$$
$$
\sum_{\rho c \gamma} [c] \pmatrix{
a&b&c\cr
\alpha&\beta&\gamma\cr
}_\rho
\pmatrix{a&b&c\cr
\alpha'&\beta'&\gamma\cr
}_\rho^*
= \delta(\alpha' \alpha) \delta(\beta' \beta)
\eqno(29)
$$
\noindent
follow by putting $R = E$ in (26) and (27). 

Relation (26) and its dual relation (27) show that ${\cal D}^a$, ${\cal D}^b$ 
and ${\cal D}^c$ present the same variance. Thus, the behaviour of the 
(3-$a \alpha)_\rho$ symbol under permutation of $a$, $b$ and $c$ should be 
easier to describe than the one of the CGc 
$(a b \alpha \beta \vert \rho c \gamma)$. This is reflected by the
integral relation (to be compared to (14)) 
$$
\vert G\vert^{-1}\int_G 
{\cal D}^a(R)_{\alpha \alpha'}
{\cal D}^b(R)_{\beta  \beta '}
{\cal D}^c(R)_{\gamma \gamma'} dR
= \sum_{\rho} \pmatrix{
a&b&c\cr
\alpha&\beta&\gamma\cr
}_\rho
\pmatrix{
a&b&c\cr
\alpha'&\beta'&\gamma'\cr
}_\rho^* 
\eqno(30)
$$
which may be proved directly by combining (27) with the great orthogonality
theorem for the group $G$. When the triple direct product 
$a \otimes b \otimes c$ contains the IRC 0 of $G$ only once (i.e., when there 
is no label $\rho$ and no summation in (30)), Eq.~(30) shows that the square 
modulus of the 3-$a \alpha$ symbol is invariant
under permutation of its columns. In this case, we may take advantage of the
arbitrariness of the matrix $M$ in (11) or (24) to produce convenient
symmetry properties of the 3-$a \alpha$ symbol 
under permutation of its columns. By way of illustration, let us mention the
following result~: For $G$ simply reducible, it is possible to
arrange that the numerical value of the 3-$a \alpha$ symbol 
be multiplied by the phase factor $(-1)^{a+b+c}$, with $(-1)^{2 x } = c_x$, 
under an odd permutation of its columns~; consequently, the numerical value of
the 3-$a \alpha$ symbol remains unchanged under an even permutation of its 
columns (since $c_a c_b c_c = 1$). 

\newpage

{\bf 4.5 Recoupling coefficients}
\smallskip 

We now define two new coefficients 
$$
(a (bc) \rho_{bc} c_{bc} \rho' d' \delta' \vert 
 (ab) \rho_{ab} c_{ab} c \rho  d  \delta) =
\sum_{\alpha \beta \gamma \gamma_{ab} \gamma_{bc}}
(a b \alpha \beta \vert \rho_{ab} c_{ab} \gamma_{ab})
(c_{ab} c \gamma_{ab} \gamma \vert \rho d \delta)
$$
$$
\times (b c \beta \gamma \vert \rho_{bc} c_{bc} \gamma_{bc})^* 
       (a c_{bc} \alpha \gamma_{bc} \vert \rho' d' \delta')^*
\eqno(31)
$$
and 
$$
((ac) \rho_{ac} c_{ac} (bd) \rho_{bd} c_{bd} \rho'e'\varepsilon'\vert
 (ab) \rho_{ab} c_{ab} (cd) \rho_{cd} c_{cd} \rho e \varepsilon)
$$
$$
=\sum_{\alpha \beta \gamma \delta}
 \sum_{\gamma_{ab} \gamma_{cd} \gamma_{ac} \gamma_{bd}}
(a b \alpha \beta  \vert \rho_{ab} c_{ab} \gamma_{ab}) 
(c d \gamma \delta \vert \rho_{cd} c_{cd} \gamma_{cd}) 
(c_{ab} c_{cd} \gamma_{ab} \gamma_{cd} \vert \rho e \varepsilon)
$$
$$
\times 
(a c \alpha \gamma \vert \rho_{ac} c_{ac} \gamma_{ac})^* 
(b d \beta \delta \vert \rho_{bd} c_{bd} \gamma_{bd})^* 
(c_{ac} c_{bd} \gamma_{ac} \gamma_{bd} \vert \rho' e' \varepsilon')^*
\eqno(32)
$$
The introduction in these definitions of (13) and the use of the great
orthogonality theorem for $G$ leads to the properties
$$
(a(bc)\rho_{bc}c_{bc} \rho' d' \delta' \vert (ab) \rho_{ab}c_{ab} c \rho d
\delta) 
$$
$$
= \delta(d' d) \delta(\delta' \delta) [d]^{-1} \sum_{\delta} 
(a(bc) \rho_{bc} c_{bc} \rho' d \delta \vert (ab) 
\rho_{ab}c_{ab} c \rho d \delta)
\eqno(33)
$$
and
$$
((ac)\rho_{ac}c_{ac} (bd)\rho_{bd}c_{bd}  \rho' e' \varepsilon' \vert 
(ab)\rho_{ab} c_{ab} (cd) \rho_{cd}c_{cd} \rho  e  \varepsilon )
$$
$$
= \delta (e'e) \delta (\varepsilon'\varepsilon) [e]^{-1} \sum_{\varepsilon}
((ac) \rho_{ac} c_{ac} (bd) \rho_{bd} c_{bd} \rho' e \varepsilon \vert 
 (ab) \rho_{ab} c_{ab} (cd) \rho_{cd} c_{cd} \rho  e \varepsilon)
\eqno(34)
$$
so that the recoupling coefficients defined by (31) and (32) are 
basis-independent (i.e., they do not depend on the labels of type $\alpha$) in
contradistinction to the coupling coefficients 
$(a b \alpha \beta \vert \rho c \gamma)$. In a way paralleling 
the passage from the coupling
coefficients to the (3-$a \alpha)_\rho$ symbol, one can define
(6-$a)_{4\rho}$ and (9-$a)_{6\rho}$ symbols 
from the recoupling coefficients defined by (31)-(34). The defining expressions 
(6-$a)_{4\rho}$ and (9-$a)_{6\rho}$ symbols are very complicated and not
especially instructive in the case of an arbitrary compact group $G$. Hence, 
they shall be omitted as well as the defining expressions for higher
$(3N$-$a)_{2N\rho}$ symbols corresponding to the recoupling of $N > 3$
IRC's. Finally, note that the recoupling coefficients and their associated 
$(3N$-$a)_{2N\rho}$ symbols,  $N > 1$,  for a group $G$ can be connected to  
other basis-independent quantities, viz., the characters of $G$. 
\medskip 

{\bf 4.6 Irreducible tensorial sets}
\smallskip 

Let $\lbrace \vert \tau a \alpha) : \alpha = 1,2,\cdots,[a] \rbrace$ 
be a basis for the irreducible matrix representation ${\cal D}^a$ of $G$. 
The vectors $\vert \tau a \alpha)$ are defined on a unitary or pre-Hilbert 
space $\cal E$ (indeed, a Hilbert space in the quantum-mechanical applications)
and there exists an application $R \mapsto U_R$ such that
$$
U_R \vert \tau a \alpha ) = \sum_{\alpha'=1}^{[a]} 
    \vert \tau a \alpha') {\cal D}^a(R)_{\alpha' \alpha}
\eqno(35)
$$
for any $R$ in $G$. The set 
$\lbrace \vert \tau a \alpha) : \alpha = 1,2,\cdots,[a] \rbrace$ 
is referred to as an irreducible tensorial set (ITS) of vectors
associated to ${\cal D}^a$. The label $\tau$ may serve to distinguish the
various ITS' of vectors associated to the same irreducible matrix
representation ${\cal D}^a$. (In practical applications, this label consists of
various quantum numbers arising from nuclear, or atomic or molecular 
configurations.) In
this connection, note the following standardization~: It is always possible to
arrange that 
$\lbrace \vert \tau a \alpha) : \alpha = 1,2,\cdots,[a] \rbrace$ and
$\lbrace \vert \tau'a \alpha) : \alpha = 1,2,\cdots,[a] \rbrace$ span the same 
matrix 
representation ${\cal D}^a$ rather than two equivalent representations.

From two ITS'
$\lbrace \vert \tau_a a \alpha) : \alpha = 1,2,\cdots,[a] \rbrace$  and 
$\lbrace \vert \tau_b b \beta ) : \beta  = 1,2,\cdots,[b] \rbrace$, we can 
construct another ITS of vectors. Let us define
$$
\vert \tau_a \tau_b a b \rho c \gamma) = \sum_{\alpha \beta} 
\vert \tau_a a \alpha) \otimes 
\vert \tau_b b \beta ) (a b \alpha \beta \vert \rho c \gamma)
\eqno(36)
$$
Then, as a simple corollary of (7), the set 
$\lbrace \vert \tau_a\tau_bab \rho c \gamma) : \gamma = 1,2,\cdots,[c] \rbrace$ 
can be shown to be an ITS associated to ${\cal D}^c$. 

In a similar way, let us consider a set 
$\lbrace T_\alpha^a : \alpha = 1,2,\cdots,[a] \rbrace$ 
of (linear) operators defined on $\cal E$ and such that
$$
U_R T_{\alpha }^a U_R^{-1} = \sum_{\alpha'=1}^{[a]} 
    T_{\alpha'}^a {\cal D}^a(R)_{\alpha'\alpha}
\eqno(37)
$$
for any $R$ in $G$. This set is called an ITS of operators associated to
${\cal D}^a$. We also say that this set defines an irreducible tensor operator
${\bf T}^a$ associated to ${\cal D}^a$. 
Note the implicit standardization~: The sets
$\lbrace T_\alpha^a : \alpha = 1,2,\cdots,[a] \rbrace$ and 
$\lbrace U_\alpha^a : \alpha = 1,2,\cdots,[a] \rbrace$ span the same matrix 
representation ${\cal D}^a$ rather than two equivalent representations.

In full analogy with (36), we define
$$
\lbrace T^a \otimes U^b\rbrace^{\rho c}_\gamma = \sum_{\alpha \beta}
T_\alpha^a U_\beta^b (a b \alpha \beta \vert \rho c \gamma)
\eqno(38)
$$
from the two ITS' 
$\lbrace T_\alpha^a : \alpha = 1,2,\cdots,[a] \rbrace$ and 
$\lbrace U_\alpha^b : \beta  = 1,2,\cdots,[b] \rbrace$. As a result, the set
$\lbrace \lbrace T^a \otimes 
                 U^b \rbrace^{\rho c}_\gamma : \gamma = 1,2,\cdots,[c] \rbrace$ 
is an ITS of operators associated to ${\cal D}^c$. We say that
$\lbrace {\bf T}^a \otimes {\bf U}^b \rbrace$ is the direct product of the 
irreducible 
tensor operators ${\bf T}^a$ and ${\bf U}^b$. 
Observe that this direct product defines a 
tensor operator which is reducible in general. Equation (38) gives the various
irreducible components of $\lbrace {\bf T}^a \otimes {\bf U}^b \rbrace$. 
\medskip

{\bf 4.7 The Wigner-Eckart theorem}
\smallskip 

The connection between most of the quantities introduced up to now appears in
the calculation of the matrix element 
$(\tau'a'\alpha'\vert T^b_{\beta} \vert \tau a \alpha)$, the scalar 
product on $\cal E$ of the vector 
                     $T^b_{\beta} \vert \tau  a  \alpha)$ by the vector 
                                 $\vert \tau' a' \alpha')$. By developing 
the identity 
$$
(\tau' a' \alpha' \vert T^b_{\beta} \vert 
 \tau  a  \alpha) = 
(\tau' a' \alpha' \vert U_R^{\dagger} U_R 
          T^b_{\beta}   U_R^{     -1} U_R \vert 
 \tau  a  \alpha)
\eqno(39)
$$
we get after some manipulations the basic theorem.

{\bf Theorem 3} (Wigner-Eckart's theorem). The scalar product 
$(\tau' a' \alpha' \vert T^b_\beta \vert \tau a \alpha)$ can be decomposed as 
$$
(\tau' a' \alpha' \vert T_\beta^b \vert \tau a \alpha) = \sum_{\rho}
( \tau' a' \vert\vert T^b \vert\vert \tau a )_\rho \; 
\sum_{ a'' \alpha'' } \pmatrix{
a''&a'\cr
\alpha''&\alpha'\cr
}
\pmatrix{
b&a&a''\cr
\beta&\alpha&\alpha''\cr
}^*_\rho
\eqno (40{\rm a})
$$
Alternatively, Eq.~(40a) can be cast into the form
$$
(\tau' a' \alpha' \vert T_\beta^b \vert \tau a \alpha) = [a']^{-\frac{1}{2}} 
\sum_{\rho } 
\langle \tau' a' \vert\vert T^b \vert\vert \tau a \rangle_\rho
 (a b \alpha \beta \vert \rho a' \alpha')^*
\eqno(40{\rm b})
$$
with 
$$
\langle \tau' a' \vert\vert T^b \vert\vert \tau a \rangle_\rho =
                              \sum_{\rho'} M (ab,a')_{\rho \rho'}^*
      ( \tau' a' \vert\vert T^b \vert\vert \tau a       )_{\rho'}
\eqno (40{\rm ab})
$$
where $M(ab,a')$ is an arbitrary unitary matrix (cf., (24) and (25)).

In the summation-factorization afforded by (40), 
there are two types of terms, namely, the (3-$a \alpha)_\rho$ symbols or 
the CGc's $(a b \alpha \beta \vert \rho a' \alpha')$ that depend on the
group $G$ only and the so-called reduced matrix elements 
$      ( \tau' a' \vert\vert T^b \vert\vert \tau a       )_\rho$ or 
$\langle \tau' a' \vert\vert T^b \vert\vert \tau a \rangle_\rho$  
that depend both on $G$ and on the physics of the problem under consideration. 
The reduced matrix elements do not depend on the `magnetic quantum numbers' 
($\alpha'$, $\beta$ and $\alpha$) and therefore, like the
recoupling coefficients, are basis-independent. We then understand the interest
of the recoupling coefficients in applications~:   The reduced matrix elements 
for a composed system may be developed as functions of reduced matrix elements 
for elementary systems and recoupling coefficients. In this direction, it can be
verified that the matrix element
$(\tau_a'\tau_b'a'b'\rho'c'\gamma'\vert\lbrace T^d \otimes U^e\rbrace{\sigma f
\atop \varphi} \vert \tau_a\tau_b a b\rho c \gamma)$ 
can be expressed in terms
of the recoupling coefficients defined by (32) and (34). 

Equations (40) generalize the Wigner-Eckart theorem originally derived
by Eckart for vector operators of the rotation group~\cite{eckart30}, 
by Wigner for tensor operators of the rotation group~\cite{wigner31} and of 
simply reducible groups~\cite{wigner65}, and 
by  Racah for tensor operators of the rotation group~\cite{racah42}. 

A useful selection rule on the matrix element 
$(\tau' a' \alpha' \vert T^b_\beta \vert \tau a \alpha)$ immediately follows
from the CGc's in (40b). The latter matrix element vanishes if the
direct product $a \otimes b$ does not contains $a'$. Consequently, in order to
have $(\tau' a' \alpha' \vert T^b_\beta \vert \tau a \alpha) \ne 0$, it is
necessary (but not sufficient in general) that the IRC $a'$ be contained in 
$a \otimes b$. 

As an interesting particular case, let us consider the situation where $b$ 
is the identity IRC 0 of $G$. This means that the operator $H=T^0_0$ is 
invariant under $G$ (see Eq.~(37)). Equation (40b) can be particularized to    
$$
(\tau' a' \alpha' \vert H \vert \tau a \alpha) = 
\delta (a' a) \delta (\alpha' \alpha) 
\langle \tau' a \vert\vert T^0 \vert\vert \tau a \rangle 
\eqno(41)
$$
where the index $\rho$ is not necessary since $a \otimes 0 = a$. The Kroneker
deltas in (41) show that there are no $a'$-$a$ and/or $\alpha'$-$\alpha$
mixing. We say that $a$ and $\alpha$ are `good quantum numbers' 
for $H$. The initial and final states have the same quantum numbers as far as
these numbers are associated to the invariance group $G$. 
The invariant $H$ does not mix state vectors belonging to different 
irreducible representations $a$ and $a'$. 
Furthermore, it does not mix state vectors 
belonging to the same irreducible representation $a$ but having different 
labels $\alpha$ and $\alpha'$. 

It is important to realize that no phase factors of the type $(-1)^a$, 
$(-1)^{a - \alpha}$ and $(-1)^{a+b+c}$ appear in (40). Indeed, the
present exposure is entirely free of
such phase factors, in contrast with other works. As a matter of fact, in many
works the passage from the Clebsh-Gordan or unsymmetrical form to the
(3-$a \alpha)_\rho$ or symmetrical form of the coupling coefficients involves
unpleasant questions of phase. This is not the case in (24) and (25). Such
a fact does not mean that (24) and (25) as well as other general
relations are free of arbitrary phase factors. In fact, all the phase
factors are implicitly contained in the matrices $M$, the 2-$a \alpha$ 
symbols and the (basis-independent) Frobenius-Schur coefficient.
\medskip

{\bf 4.8 The Racah lemma}
\smallskip 

We have already emphasized the interest of considering chains of groups
rather than isolated groups. Let $K$ be 
a subgroup of $G$. In this case, the labels
of type $\alpha$ that occur in what preceeds, may be replaced by triplets of
type $\alpha \Gamma \gamma$. The label of type $\Gamma$ stands for an IRC of
the group $K$, the label of type $\gamma$ is absolutely necessary when 
$[{\Gamma}] > 1$ and the label
of type $\alpha$ is now a branching multiplicity label to be used when the IRC
$\Gamma$ of $K$ is contained several times in the IRC $a$ of the head group
$G$. 
(The label $\gamma$ is an internal multiplicity label for $K$ and the label 
$a$ is an external multiplicity label inherent to the restriction $G \to K$.) 
Then, the CGc $( a_1 a_2 \alpha_1 \alpha_2 \vert \rho a \alpha )$ for the group
$G$ is replaced by  the CGc 
 $( a_1 a_2 \alpha_1 \Gamma_1 \gamma_1 
            \alpha_2 \Gamma_2 \gamma_2 \vert \rho a \alpha \Gamma \gamma )$ for
the group $G$ in a $G \supset K$ basis. We can prove the following theorem. 

{\bf Theorem 4} (Racah's lemma). The CGc's of the group $G$ 
in a $G \supset K$ basis can be developed according to 
$$
( a_1 a_2 \alpha_1 \Gamma_1 \gamma_1 
          \alpha_2 \Gamma_2 \gamma_2 \vert \rho a \alpha \Gamma \gamma ) =
\sum_{\beta} 
( \Gamma_1 \Gamma_2 \gamma_1 \gamma_2 \vert \beta \Gamma \gamma )
(a_1 \alpha_1 \Gamma_1 + 
 a_2 \alpha_2 \Gamma_2 \vert \rho a \alpha \Gamma)_\beta 
\eqno (40{\rm c})
$$
where the coefficients 
$( \Gamma_1 \Gamma_2 \gamma_1 \gamma_2 \vert \beta \Gamma \gamma )$ are CGc's
for the group $K$ as considered as an isolated group and the coefficients 
$(a_1 \alpha_1 \Gamma_1 + 
 a_2 \alpha_2 \Gamma_2 \vert \rho a \alpha \Gamma)_\beta$ do not depend on 
$\gamma_1$, $\gamma_2$ and $\gamma$. 

The proof of Racah's lemma was originally 
obtained from Schur's lemma~\cite{racah42}. However, the 
analogy between (40a), (40b) and (40c) should be noted. Hence, the Racah lemma
for a chain $G \supset K$ may be derived from the Wigner-Eckart theorem, 
for the group $G$ 
in a $G \supset K$ basis, applied to the Wigner operator, i.e., the operator 
whose matrix elements are the CGc's. The expansion coefficients       
$(a_1 \alpha_1 \Gamma_1 + 
 a_2 \alpha_2 \Gamma_2 \vert \rho a \alpha \Gamma)_\beta$ 
in the development (40c) are sometimes
named isoscalar factors, a terminology that comes from the 
chain ${\rm SU}(3) \supset {\rm SU}(2)$ used in the eightfold way model 
of subatomic physics. 

From a purely group-theoretical point of view, it is worth to note that Racah's
lemma enables us to calculate the CGc's of the subgroup $K$ of $G$ 
when the ones of the
group $G$ are known~\cite{kibler7982}. In particular, for those triplets 
$(\Gamma_1\Gamma_2\Gamma)$ for which $\Gamma_1 \otimes \Gamma_2$ contains
$\Gamma$ only once, the CGc's
$( \Gamma_1 \Gamma_2 \gamma_1 \gamma_2 \vert \Gamma \gamma )$ are given by a
simple formula in terms of the CGc's of $G$. 

The summation-factorisation in (40c) can be applied to each CGc
entering the definition of any recoupling coefficient for the group $G$. 
Therefore, the recoupling coefficients for $G$ can be developed in terms of the
recoupling coefficients for its subgroup $K$~\cite{kibler7982}.
\medskip 

{\bf 4.9 The rotation group}
\smallskip 

As an illustrative example, we now consider 
the universal covering group or, in the terminology of
molecular physics, the `doubled' group SU(2) of the proper 
rotation group $R(3)$. In
this case, $a \equiv j$ is either an integer (for vector representations) or a
half-an-odd integer (for spinor representations), $\alpha \equiv m$ ranges from
$-j$ to $j$ by unit step, and ${\cal D}^a(R)_{\alpha\alpha'}$ identifies to the
element ${\cal D}^{(j)}(R)_{mm'}$ of the well-known Wigner rotation matrix of
dimension $[j] \equiv 2j +1$. The matrix representation ${\cal D}^{(j)}$
corresponds to the standard basis 
$\lbrace\vert jm) : m = -j, -j + 1, \cdots, j\rbrace$ where
$\vert jm)$ denotes an eigenvector of the (generalized) 
angular momentum operators $J^2$ and
$J_z$. (For $j$ integer, the label $\ell$ often replaces $j$.) The labels of 
type $m$
clearly refer to IRC's of the rotation group $C_{\infty} \sim R(2)$. 
Therefore, the basis 
$\lbrace\vert jm) : m = -j, -j + 1, \cdots, j\rbrace$
is called and $R(3) \supset R(2)$ or ${\rm SU}(2) \supset {\rm U}(1)$ basis. 
Furthermore, the
multiplicity label $\rho$ is not necessary since SU(2) is multiplicity-free.
Consequently, the (real) CGc's of SU(2) in a ${\rm SU}(2) \supset {\rm U}(1)$ 
basis are written $(j_1j_2m_1m_2\vert jm)$. They are also called Wigner 
coefficients. 

In view of the ambivalent nature of SU(2), the 2-a$\alpha$ symbol reduces to 
$$
\pmatrix{
j&j'\cr
m&m'\cr
} = (-1)^{j+m} \delta(j'j) \delta (m',-m)
\eqno(42)
$$
where $(-1)^{j+m} \delta (m',-m)$ corresponds to the 
1$-jm$ Herring-Wigner metric tensor. Then, the
introduction of (42) into (24) for the chain 
${\rm SU}(2) \supset {\rm U}(1)$ shows
that the 3-a$\alpha$ symbol identifies to the 3-$jm$ Wigner symbol
$$
\pmatrix{
j_1&j_2&j_3\cr
m_1&m_2&m_3\cr
} = (2j_3 + 1)^{-\frac{1}{2}} (-1)^{j_3 - m_3 - 2 j_2} 
( j_2 j_1 m_2 m_1 \vert j_3, - m_3 )
\eqno (43) 
$$ 
provided we chose $M(j_2j_1,j_3) = (-1)^{2j_1}$. Such a choice ensures that 
the 3-$jm$ symbol is highly symmetrical under permutation of its columns.

In the SU(2) case, the (6-a)$_{4\rho}$ and (9-a)$_{6\rho}$ 
symbols may be chosen to coincide  with  the 6-$j$ Wigner 
(or $\bar W$ Fano-Racah) symbol      and the 9-$j$ Wigner 
(or $     X$ Fano-Racah) symbol, respectively. More precisely, we may take 
$$
\left\lbrace\matrix{
j_1&j_{23}&j  \cr 
j_3&j_{12}&j_2\cr
}\right\rbrace
= (-1)^{j_1+j_2+j_3+j} [(2j_{12}+1) (2j_{23}+1)]^{-\frac{1}{2}}
$$
$$
\times (j_1(j_2j_3)j_{23}jm \vert (j_1j_2)j_{12}j_3jm)
\eqno (44)
$$
and 
$$
\left\lbrace\matrix{
j_1   &j_2   &j_{12}\cr
j_3   &j_4   &j_{34}\cr
j_{13}&j_{24}&j     \cr
}\right\rbrace
= [(2j_{12} + 1)(2j_{34} + 1)(2j_{13} + 1)(2j_{24} + 1)]^{-\frac{1}{2}}
$$
$$
\times ((j_1j_3)j_{13}(j_2j_4)j_{24}jm \vert (j_1j_2)j_{12}(j_3j_4)j_{34}jm)
\eqno (45)
$$
in terms of recoupling coefficients (cf., (33) and (34)).

Finally, for $a \equiv k$, the ITS {\bf T}$^a$ coincides with the
irreducible tensor operator {\bf T}$^{(k)}$ 
of rank $k$ (and having $2 k + 1$ components) 
introduced by Racah. 
We denote by $T^{(k)}_q$ the ${\rm SU}(2) \supset {\rm U}(1)$ components 
of {\bf T}$^{(k)}$.

All the relations of subsections 4.1-4.7 
may be rewritten as familiar relations of angular 
momentum theory owing to the just described correspondence 
rules. For example, Eqs.~(17) or (18) and (40a) can 
be specialized to 
$$
{\cal D}^{(j)}(R)_{mm'}^* = (-1)^{m-m'} {\cal D}^{(j)}(R)_{-m,-m'}
\eqno (46)
$$
and
$$
(\tau' j' m' \vert T^{(k)}_q \vert \tau j m) = (-1)^{j'-m'} \pmatrix{
j'&k&j\cr
-m'&q&m\cr
}
(\tau' j' \vert\vert T^{(k)} \vert\vert \tau j)
\eqno (47)
$$
respectively.
\bigskip

\noindent {\bf 5~~Applications} 
\medskip

There exists a huge literature on the application of symmetries considerations 
to physics. See for instance Refs.~[3,7,11] for atomic physics, 
Refs.~[8-10,13] for molecular and condensed matter physics and Refs.~[12,13]
for nuclear physics. It is not feasible to give here a detailed account of all 
possible applications. Thus, we limit ourselves to a list of important points 
for dealing with the applications. 

{\bf Degeneracies}. The minimal degeneracies for the eigenvalues of an operator
$H$ invariant under a group $G$ can be predicted prior to any calculation. They
correspond to the dimensions of the irreducible representations of $G$. The 
explanation of the accidental (with respect to $G$) degeneracies of $H$, if
any, lies on the existence of a larger invariance group.

{\bf State labelling}. The eigenvalues (e.g., energy levels or masses) of an
operator (e.g., Hamiltonian or mass operator) $H$ invariant under a group 
$G$ can be classified with the help of the irreducible representations of a 
chain of groups involving the symmetry or invariance group $G$. The 
eigenvectors (state vectors or wavefunctions) of $H$ can be also classified 
or labelled with irreducible representations of the groups of the chain. 

{\bf Operator labelling}. For a given physical system, the interactions
(involving the Hamiltonian) can be classified according to their transformation
properties under a chain of groups. Such a classification is a pre-requisite for
the application of the Wigner-Eckart theorem. The queue group in the chain
generally corresponds to an actual (or idealized) symmetry and the other groups
to approximate symmetries and/or classification groups. If the interactions are
known the chain can be derived by inspection. On the other hand, if the
interactions are postulated, the chain follows from physical and mathematical
arguments. In this case, the Hamiltonian for the system is written as a linear
combination of operators invariant under the various groups of the chain. The
coefficients of the linear combination are phenomenological parameters to be 
fitted to experimental data and interpreted in the framework of (microscopic) 
models. Along this vein, we can quote the Iachello-Levine vibron model based on
the chains 
${\rm U}(4) \supset {\rm SO}(4) \supset {\rm SO}(3) \supset {\rm SO}(2)$ 
and
${\rm U}(4) \supset {\rm  U}(3) \supset {\rm SO}(3) \supset {\rm SO}(2)$ 
for diatomic molecules [13]. Let us also mention the Arima-Iachello interacting
boson model (IBM) and the interacting boson-fermion model in nuclear structure
physics [12,13]. The three chains  
${\rm U}(6) \supset {\rm  U}(5) \supset {\rm SO}(5) \supset {\rm SO}(3)$, 
${\rm U}(6) \supset {\rm SO}(6) \supset {\rm SO}(5) \supset {\rm SO}(3)$ 
and
${\rm U}(6) \supset {\rm SU}(3) \supset {\rm SO}(3)$ are used in the 
IBM~; they correspond to three different regimes. 

{\bf Level splitting}. The splitting of the spectrum of an operator $H_0$
invariant under a group $G_0$ when we pass, in a perturbative or
nonperturbative way, from $H_0$ to $H_0 + H_1$, where $H_1$ is invariant under
a subgroup $G_1$ of $G_0$, can be determined by studying the restriction 
$G_0 \to G_1$. Familiar examples concern the Zeeman effect, the
(homogeneous) Stark effect, and the crystal- and ligand-field 
effects.\footnote{\small Crystal- and ligand-field theories are 
standard tools for analyzing thermal, optical and magnetic properties 
of transition elements ($nd^N$ and $nf^N$ ions) in crystals [8-10].} 

More quantitative applications concern~: (i) The effective determination of
symmetry adapted vectors (e.g., normal vibration modes of a molecule, symmetry
adapted functions like molecular orbitals in the framework of the 
linear combination of atomic orbitals (LCAO) method, 
$N$-particle wavefunctions)~; (ii) The factorization of the secular equation~;
(iii) The determination of selection rules. We briefly discuss in turn these
points. 

{\bf Normal modes}. The determination of the normal vibration modes of a
molecule or complex ion or small aggregate goes back (with the theory of level
splitting for ions in crystals) to the end of the twenties. It is based on the
reduction of a representation arising from the transformation properties of the
molecular skeleton. 

{\bf Symmetry adapted functions}. In general, functions (e.g., atomic
and nuclear wavefunctions, molecular orbitals, etc.) can be obtained from the
method of projection operators developed by, among others, Wigner, L\"owdin and
Shapiro. This is sometimes referred to as the Van Vleck generating machine~:
The action on an arbitrary function 
of a projection operator associated to an irreducible representation
$\Gamma$ of a group $G$ produces nothing or a function transforming according to
$\Gamma$. One then easily understands why the atomic orbitals occurring in a
LCAO molecular orbital have the same symmetry. This fact illustrates a general
characteristic when dealing with symmetries~: A mixing between state vectors of
different symmetries with respect to a group $G$ can be performed only 
owing to a transition operator which is not invariant under $G$. 

{\bf $N$-particle wavefunctions}. The concepts of seniority and coefficients of
fractional parentage (cfp's) introduced by Racah [7] provides us with an
alternative to the determinental Slater method.   The wavefunctions for a system
of $N$ equivalent particles (nucleons or electrons) can be developed in terms
of the wavefunctions for a subsystem of $N-1$ particles. The expansion
coefficients involve cfp's which can be thought of Clebsch-Gordan 
coefficients for a chain of
groups. These Clebsch-Gordan 
coefficients can be factorized according to Racah's lemma. 

{\bf Secular equation}. A cornerstone for the application of group theory to
physics is the Wigner-Eckart theorem in its generalized form that allows one to
calculate matrix elements and to 
build secular equations (energy matrices or mass
matrices). Since it is not possible to admix, via an operator invariant 
$H$ under a
group $G$, wavefunctions belonging to different irreducible representations of
$G$ (cf., Eq.~(41)), a secular equation for $H$ can be factorized into
blocks corresponding to distinct irreducible representations of $G$. 

{\bf Selection rules}. Such rules already occur in the determination of matrix
elements. The are particularly useful in the determination of intensities
$| \sum (f \Gamma_f \gamma_f |T^{\Gamma}_{\gamma} | i \Gamma_i \gamma_i ) |^2$ 
for the transitions
induced by a tensor operator ${\bf T}^{\Gamma}$ between an initial state $i$
and a final state $f$.  

To summarize this paper, we have shown how symmetries occur in nuclear, atomic,
molecular and (to some extent) condensed matter physics through the introduction
of groups and chains of groups. Such chains involve symmetry and
classification groups. Group theory is thus an important tool for
classification purposes by means of `boxes' constituted by irreducible
representations of groups. The classification may concern wavefunctions and
interactions.\footnote{\small To be complete, the classification may also
concern atomic and subatomic particles (cf., the group SU(3) for the three 
light 
flavors of quark and the group SU(2)$\otimes$SO(4,2) for the periodic table of
chemical elements) as well as fields (e.g., the matter and gauge fields of the 
SU(3)$_c \otimes$SU(2)$_L \otimes$U(1)$_Y$ $\supset$ 
SU(3)$_c \otimes$U(1)$_Q$ standard model of particle physics).} We have also
examined how the Wigner-Racah algebra associated to a chain of groups provides
us with a useful tool for calculating quantum-mechanical matrix elements. To
close, let us mention that other structures (more specifically, 
graded Lie groups, 
graded Lie algebras, 
quantum groups and Hopf algebras) are in the present days the object of an
intense activity in connection with `new symmetries' (as, for example,
supersymmetries). 
\bigskip

{\bf Appendix~: Note on the Hydrogen Atom}
\medskip
  
Let us start with the Hamiltonian  $H$  for a three-dimensional hydrogenlike 
atom (see Eq.~(0) of Example 9). 
(Equation (0)) follows after separation of the collective 
and electronic motions. The units are such that $e = \hbar = 1$ and the 
reduced mass electron-nucleus is taken to be equal to 1.) In Eq.~(0), we take
$$
\Delta = \sum_{a=1}^3 \frac{\partial^2}{\partial x_a^2} \qquad 
r = \sqrt{\sum_{a=1}^3 x_a^2}
$$
The components $L_a$ with $a=1,2,3$ of the angular 
momentum of the electron can be written as
$$
L_a = \frac{1}{2} \varepsilon_{abc} L_{bc}    \qquad {\rm where} \qquad
                                     L_{ab} = x_a p_b - x_b p_a \qquad 
{\rm and} \qquad p_a = - {\rm i} \frac{\partial}{\partial x_a} 
$$
We know that the observables $L_a$ with $a=1,2,3$ and 
$L^2 = \sum_{a=1}^3 L_a^2$ are constants of motion since 
$$
[L^2, H] = [L_a, H] = 0 \qquad a = 1,2,3
$$
Other constants of the motion are provided by the Runge-Lenz-Pauli 
vector\footnote{\small It is the quantum-mechanical analogue of 
the Laplace-Runge-Lenz vector well known in classical mechanics.} 
${\vec M} (M_1,M_2,M_3)$ defined by 
$$
\vec M = - Z \frac{\vec r}{r} + \frac{1}{2}
(\vec p \times \vec L - \vec L \times \vec p )
$$
in terms of the vectors ${\vec p} (p_1,p_2,p_3)$ and  ${\vec L} (L_1,L_2,L_3)$.
We can check that
$$
[M^2, H] = [M_a, H] = 0 \qquad a = 1,2,3
$$
where $M^2 = \sum_{a=1}^3 M_a^2$.

We now ask the question~: What becomes the Lie algebra so(3) (of the group
SO(3)), spanned by the operators $L_a$, when we introduce the operators
$M_a$~? In this respect, we have the following commutation relations 
$$
[L_a, L_b] =    {\rm i}         \varepsilon_{abc} L_c   \qquad
[L_a, M_b] =    {\rm i}         \varepsilon_{abc} M_c   \qquad
[M_a, M_b] =    {\rm i} (- 2 H) \varepsilon_{abc} L_c 
\eqno ({\rm A.}1)
$$
Equations (A.1) do not define a Lie algebra due to the presence of the operator 
$H$ that prevents $[M_a, M_b]$ to be a linear combination of the $L_a$'s and 
$M_a$'s. In addition to (A.1), we can show that
$$
\sum_{a=1}^3 L_a M_a = \sum_{a=1}^3 M_a L_a = 0 \qquad 
M^2 - Z^2 = 2 H (L^2 + 1)
\eqno ({\rm A.}2)
$$
From Eqs.~(A.1) and (A.2), it is possible to construct an infinite dimensional 
Lie algebra or even a finite $W$ algebra. We shall not do it here. We shall
rather consider (A.1) for the various parts (discrete, continuous and 
zero-energy point) of the spectrum of $H$. 

{\bf 1}. For the discrete spectrum, we have $H < 0$ in (A.1) and we
introduce 
$$
A_a = \sqrt{- \frac{1}{2H}} M_a  \qquad
J_a = \frac{1}{2} (L_a + A_a)    \qquad
K_a = \frac{1}{2} (L_a - A_a)    \qquad a = 1, 2, 3
$$
This leads to
$$
[J_a, J_b] = {\rm i} \varepsilon_{abc} J_c \qquad
[K_a, K_b] = {\rm i} \varepsilon_{abc} K_c \qquad
[J_a, K_b] = 0
$$
which is reminiscent of the six-dimensional 
Lie algebra so(4) of the group SO(4) in an 
${\rm so}(3) \oplus {\rm so}(3)$ basis. It is clear that the generators 
of SO(4) commute with $H$, a fact that ensures that $H$ is invariant under 
SO(4). 

The discrete energy spectrum then follows from (A.2) and the 
quantization of $\{ J^2 = \sum_{a=1}^3 J_a^2, J_3 \}$ and 
                $\{ K^2 = \sum_{a=1}^3 K_a^2, K_3 \}$. The first constraint 
relation (A.2) yields $J^2 = K^2$. Let $j(j+1)$ with $2j \in {\bf N}$ be the 
common eigenvalues of $J^2$ and $K^2$. Then, by introducing $J^2=K^2=j(j+1)$ 
and $H=E$ in the second constraint relation (A.2), we arrive at the familiar 
result
$$
E \equiv E_n = - \frac{1}{2} \frac{Z^2}{n^2} \qquad n = 2j+1 \in N^*
$$
The degree of degeneracy of $E_n$, namely $(2j+1)(2j+1) = n^2$, is obtained by
counting the number of states arising from a subspace of eigenvectors, 
associated to the quantum number $j$, of the two commuting sets of operators 
$\{ J^2, J_3 \}$ and $\{ K^2, K_3 \}$. In conclusion, the level $E_n$ is 
linked to the irreducible representation $(j,j)$, with $j = (n-1)/2$, of SO(4). 

{\bf 2}. For the continuous spectrum, we have $H > 0$ in (A.1) and we put 
$$
B_a = \sqrt{ \frac{1}{2H} } M_a  \qquad a = 1, 2, 3
$$
This leads to
$$
[L_a, L_b] =   {\rm i} \varepsilon_{abc} L_c \qquad
[L_a, B_b] =   {\rm i} \varepsilon_{abc} B_c \qquad
[B_a, B_b] = - {\rm i} \varepsilon_{abc} L_c 
$$
which corresponds to the Lie algebra so(3,1) of the group SO(3,1). 

{\bf 3}. For the zero-energy point, we have $H = 0$ in (A.1). We thus
obtain 
$$
[L_a, L_b] = {\rm i} \varepsilon_{abc} L_c \qquad
[L_a, M_b] = {\rm i} \varepsilon_{abc} M_c \qquad
[M_a, M_b] = 0
$$
which defines the Lie algebra e(3) of the Euclidean group E(3).

\end{document}